\documentclass[12pt]{iopart}
\usepackage{graphicx}

\begin{document}
\title[]{Influence of finite baryon density on hadronization in nucleus-nucleus
collisions via recombination}
\author{C.B. Yang
\footnote[3]{To whom correspondence should be addressed (cbyang@mail.ccnu.edu.cn)}
and H. Zheng}
\address{Institute of Particle Physics, Central China Normal University,
Wuhan 430079, P.R. China}

\begin{abstract}
In this paper is investigated the influence of net baryon density on baryon
and meson yields in relativistic nucleus-nucleus collisions, based on the recombination
model for hadronization. Unitarity condition is used as a constraint on the model. Three
cases with different assumptions on the expansion of partonic system
are considered and the baryon to meson ratio is calculated for those
situations.
\end{abstract}


\maketitle

\section{Introduction}
Partons, both soft and hard, are produced in high energy nucleus-nucleus collisions.
Because of color confinement colored objects (quarks and gluons) must convert to some
hadrons which can be detected experimentally. The problem we concern in this paper is
how those partons convert into the observed final state hadrons and the relation of
hadron yield with the parton density. On the topics of hadron production, there are
two widely used traditional models: the string model for soft hadron
production and the Feynman and Field's independent fragmentation model for hard
hadron production. The string model worked well for elementary collisions, such as
$e^+e^-$ annihilations, where strings can be formed among a few initial partons
and break up to form the final state hadrons.  In relativistic heavy ion collisions,
there are thousands initial partons. It is very hard to pair partons and have a string
for each pair. Even if strings are formed, their properties must be modified by the
presence of many other color charges. So the soft hadron production must be different from
that in elementary collisions. In some researches such a modification is treated by
the fusion of strings in terms of percolation model \cite{string}.
The fragmentation approach, on the other hand, is based
on the factorization theorem which is true only for processes with large momentum transfer
and predicts a small baryon over meson ratio. This prediction is in contradiction with
experimental observations at RHIC in Au+Au collisions, where the $p/\pi$ ratio can exceed one
at $p_T\sim 3$ GeV$/c$ \cite{p_pi}, showing strong influence of the hot medium
produced in the collisions. RHIC also observed the constituent quark number scaling
for the elliptic flow \cite{flow}, species dependence of the Cronin effect \cite{cronin}, and
jet shape modifications \cite{asso}. All these observations cannot be explained with the two
traditional hadronization models.

Last a few years witnessed the rapid development of the quark recombination models
\cite{reco,mv,anom,fb} as
a new approach for the problem of hadronization in relativistic heavy ion collisions
at RHIC. Viewed in these models, the colliding system generates partons which evolve
in phase space according to the principles of quantum chromodynamics. In this process,
quarks get dressed gradually. Finally, when the energy density (or temperature) reaches the
critical value, the parton system hadronizes into the final state hadrons.
The quark recombination model deals with only the final stage, i.e. the process from
dressed quarks to the hadrons. A basic assumption in all the recombination models is that
the final state hadrons are formed by recombining two (or three) dressed (anti)quarks.
An important feature for the models is the efficiency
in producing hadrons with intermediate transverse momentum from thermal (or soft)
parton system. The efficiency comes from the fact that the momentum of a hadron is
the sum of that of the constituent quarks forming the hadron. Because of the abundance
of soft quarks, the recombination approach ensures high yield of mesons with intermediate
transverse momentum. More importantly, the interactions among soft quarks and shower partons
in jets are the key in explaining experimental
observations such as the constituent quark number scaling of the elliptic flow, the
modification of the jet structure, the species dependence of the Cronin effect etc.,
because of the colliding system dependence of the produced (hot) soft medium.

There are, however, some difficulties with this model. The yield of meson (baryon)
is proportional to the square (cubic) of quark density or constituent quark number.
This dependence violates unitarity, because the number of mesons (baryons) produced will
increase by a factor of four (eight) if the quark density increases by a factor of two.
As a result the numbers of dressed constituent quarks can not be conserved
and unitarity is violated. A way to overcome this difficulty is to reformulate
the quark recombination as a dynamical process with the introduction of hadronization
time \cite{yang1}. At each moment, the production rates for mesons and baryons are
proportional to square and cubic of the quark density, so the basic idea in the recombination
model retains. The total yields are, however, constrained by the conservation of numbers of
constituent quarks, and the unitarity condition is also satisfied in this approach.

In \cite{yang1} only the simplest case with zero net baryon density was considered. In that
case, the number of baryons produced is equal to that of anti-baryons. So one
can basically consider only the production of mesons and baryons. In real Au+Au collisions
the net baryon density in the central rapidity region is not zero because of the nuclear
stopping effect. The net baryon density enables more baryons produced than anti-baryons. It
would be important to know how the ratio of baryon to meson depends on the net baryon
density. This is the issue we want to discuss in this paper.
We only investigate the yields of mesons and baryons but not the spectra for the final
state hadrons, as done in earlier work of the coalescence model \cite{coals}.

\section{Basic formulism}
In every implementation of the quark recombination models, the transverse
momentum spectrum of mesons at mid-rapidity can be written, after some algebras, as
\begin{equation}
\frac{dN^M}{p_Tdp_T}=\int dp_1dp_2F(p_1,p_2)R^M(p_1,p_2,p_T)\ ,
\label{old}
\end{equation}
where a factor $\delta(p_T-p_1-p_2)$ is included in the recombination
function $R^M(p_1,p_2,p_T)$ to ensure momentum conservation and $F(p_1,p_2)$
is associated with the joint quark-antiquark momentum distribution.
For baryon production a similar equation can be written out.
When we are interested in the total
multiplicity of a kind of hadron, we can consider the contribution from
recombination of pure thermal partons only, because most produced hadrons are in the
low transverse momentum region where pure thermal recombination
dominates. In the region of low transverse momentum, the joint distribution for
quark-antiquark pair can be written as $F(p_1,p_2)=V\rho^2f_1(p_1)f_2(p_2)$
with $V$ the spatial volume of the partonic system and $\rho$ the thermal
parton density just before hadronization when the (anti)quark transverse momentum
distributions $f_{1,2}$ are normalized to some fixed constant. Then one can carry out the
integration over $p_T$ and the result shows that the yield of meson (baryon)
is proportional to the square (cubic) of quark density at hadronization.
This shows a violation of unitarity, since the total number of constituent quarks
should be conserved because of their low virtuality and energy and the total
number of hadrons produced must be about proportional to the total quark number.
In \cite{yang1} it was suggested that the production rate for meson (baryon)
is proportional to the square (cubic) of the quark density, but the hadronization
process conserves the total constituent quark numbers. In this way, the main features
of the quark recombination models retain and unitarity is also conserved.
Because we are going to consider the case with nonzero net baryon density,
the yields of baryon and anti-baryon are different. Thus different from consideration
in \cite{yang1} one should have anti-baryon yield as an additional variable
being taken into account. For simplicity we consider hadronization of a partonic
system with light quarks ($u, d, \bar{u}$ and $\bar{d}$) and catalog final state hadrons
by their baryon numbers only. Extension to include strange quarks is straight forward.
 The rate equations for the
yields of baryons and mesons can be expressed as
\begin{eqnarray}
&{dN_M\over dt} &=A_MV\rho_q(t)\rho_{\bar{q}}(t)\ ,\nonumber\\
&\frac{dN_B}{dt} & = A_BV\rho_q^3(t)\ ,\label{basic}\\
&\frac{dN_{\bar{B}}}{dt} & = A_BV\rho_{\bar{q}}^3(t)\ ,\nonumber
\end{eqnarray}
with $V$ the volume of partonic system, $\rho_q$ and $\rho_{\bar{q}}$ densities for
quarks and antiquarks. In last equations, $A_B$ and $A_M$ are determined mainly by
the hadronic structures of meson and baryon. The information of both the shape of
quark distributions and the recombination functions is encoded in $A$'s. From the
reaction-rate theory \cite{react} $A_M$ and $A_B$ are also proportional to the
corresponding cross-sections and the degeneracy factors. Normally $A_B/A_M\ll 1$, and
densities of quarks and antiquarks are a few times higher than that in normal nuclear matter.
 In quark recombination models gluons are assumed to have turned into $q\bar{q}$
before hadronization. So there is no term for gluon contribution in
last equations. Because hadronization takes place at low temperature, collisions
between low momentum quarks and formed hadrons will not likely break
the hadrons into quarks. In addition most of the produced hadrons will fly
out of the reaction zone and will not collide with quarks left. As a reflection
of this fact, there is no reverse term in last equations
for the process $q+{\rm hadrons}\to {\rm quarks}$. The same assumptions
are adopted in \cite{yang1}.
The conservation conditions for the numbers of constituent quarks read
\begin{eqnarray}
&\frac{d(V\rho_q)}{dt} & =-3\frac{dN_B}{dt}-\frac{dN_M}{dt}\ ,\label{yield}\\
&\frac{d(V\rho_{\bar{q}})}{dt} & =-3\frac{dN_{\bar{B}}}{dt}-\frac{dN_M}{dt}\ . \nonumber
\end{eqnarray}
The last two expressions in last equations ensure
the unitarity in hadronization process. The initial conditions are:
$V(t=0)=V_0,\ \rho_q(t=0)=\rho_0, \rho_{\bar{q}}(t=0)=\kappa \rho_0$, with $V_0, \rho_0,
\kappa$ parameters to be input from other models. Eqs. (2) and (3) form the fundamental
formulas for the yields of mesons and baryons in hadronization from quark recombination
model with unitarity constraint.

One can see easily that the above equations are not closed. To solve these equations
additional input on the relation of $\rho_q, \rho_{\bar{q}}$ and $V$  must be introduced
from elsewhere. In ultra-relativistic heavy ion collisions
the produced partonic system expands. If the expansion of the parton system retains
to the last stage of its evolution and the transition from quarks to hadrons is of first-order,
there are two competing trends in hadronization on the change of the volume. One trend is
the system's hydrodynamical expansion which will make the partonic volume larger,
and the other is hadronization process which happens on the surface of the system and
tries to shrink the system. Thus the hadronization dynamics should, in general, be
connected with the hydrodynamical calculations. Such combination is beyond the scope
of this paper. Even without hydrodynamical input, qualitative behaviors of $N_B$ etc can
be made from Eq. (2). The behaviors of the density $\rho$'s for all the situations discussed
below are very similar to those shown in \cite{yang1} for the corresponding cases,
and we will not show them in this paper because they are not observable.
Since $\rho_q$'s decrease with time in hadronization, production
rates for baryons and mesons also decrease with time, more obvious for baryons than for
mesons. So $N_B, N_M$ etc increase with time rapidly at first, then slow down, and
finally saturate at the end of hadronization.

In this paper we only discuss the situations we can do numerically
without involving hydrodynamical calculations. We will show some results
under simplest assumptions. This is enough for illustrating the net baryon density dependence
in hadronization from the quark recombination model.

\section{Main results}
\subsection{For fixed volume of the system}
  The first case one can investigate is with fixed $V=V_0$. For later
convenience, we define
\begin{eqnarray*}
n_B\equiv N_B/(\rho_0 V_0)\ ,
n_M\equiv N_M/(\rho_0 V_0)\ ,
n_{\bar{B}}\equiv N_{\bar{B}}/(\rho_0 V_0)\ .
\end{eqnarray*}
They are the average hadron multiplicities produced from per constituent quark
in the state just before hadronization. After
integrating Eq. (3) over $t$ one gets
\begin{eqnarray*}
3n_B+n_M&=&1\ ,\\
3n_{\bar{B}}+n_M&=&\kappa\ .
\end{eqnarray*}
So $n_B-n_{\bar{B}}=(1-\kappa)/3$, independent of
model assumptions in this paper. In fact this expression is nothing more than the
conservation of net baryon number in hadronization.
We also define $r=\rho_{\bar{q}}/\rho_q, u=A_B\rho_0/A_M, \rho=\rho_q/\rho_0$ and
$\tau=A_M\rho_0 t$, then Eqs. (\ref{basic}) and (\ref{yield}) can be
rewritten as
\begin{eqnarray}
\frac{d\rho}{d\tau}& =& -3u \rho^3-r\rho^2\ ,\nonumber\\
\frac{dr}{d\tau}&=& 3u\rho^2(r-r^3)-\rho(r-r^2)\ ,\nonumber\\
\frac{dn_B}{d\tau}&=& u\rho^3\ ,\\
\frac{dn_M}{d\tau}&=& r\rho^2\ ,\nonumber\\
\frac{dn_{\bar{B}}}{d\tau}&=& ur^3\rho^3\ .\nonumber
\end{eqnarray}
Initial conditions for last equations are $\rho(0)=1, r(0)=\kappa$.
The obtained yields $n_B$ etc depend on values of parameters $u$ and $\kappa$.
We try to input two typical values for $\kappa=0.8$ and 0.6 and investigate
the yields as functions of $u$ which depends on the initial quark density $\rho_0$
and the competition factor $A_B/A_M$ of baryon production relative to that of meson.
A large value of $u$ may be caused due to a higher initial quark density or larger probability
for baryon production relative to that for mesons. The obtained results for the yields
are shown in Fig. 1.
\begin{figure}[tbph]
\centerline{\includegraphics[width=0.5\textwidth]{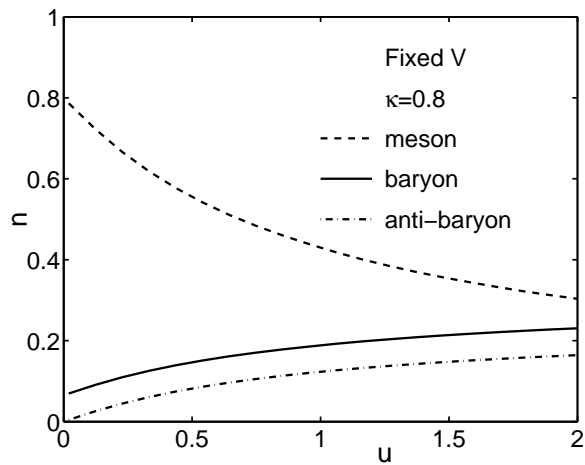}}
\caption{Yields for baryon, meson, and anti-baryon for the case with fixed system
volume as functions of $u$ for given $\kappa=0.8$.}
\end{figure}
From the figure one can see that with the increase of $u$ more baryons but less mesons
can be produced. This is from the different dependence of their production rates
on the quark density. The enhancement of baryon production can be seen more
clearly from the ratio $R_{B/M}=n_B/n_M$ as a function of $u$, as shown in Fig. 2.
The baryon to meson ratio increases almost linearly with $u$ and can be larger than
0.8 at $u=2$ for $\kappa=0.6$. One may have noticed that the ratio is larger at the same $u$
when the parameter $\kappa$ is smaller. This is not surprising, because if $\kappa$ is smaller
there are less anti-quarks and thus less mesons can be produced from the system
while most of the quarks can only recombine to form baryons.

\begin{figure}[tbph]
\centerline{\includegraphics[width=0.5\textwidth]{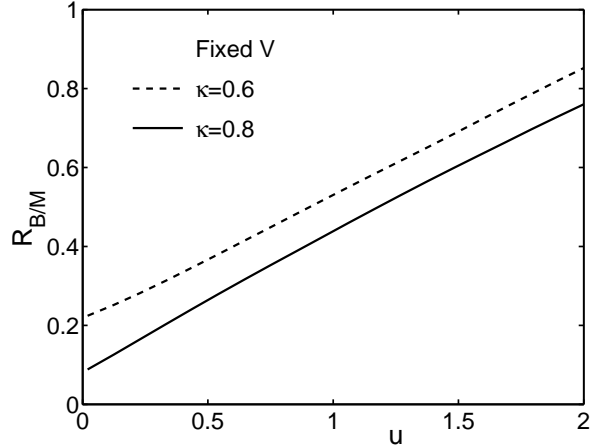}}
\caption{Ratio of baryon yield to that of meson as a function of $u$
for given $\kappa=0.8$ and 0.6 for the first case.}
\end{figure}

\subsection{For fixed quark density}
The second case we can consider is when the quark density is assumed to remain
unchanged in hadronization. Then the volume of the system and the anti-quark density
can change in the process. We define a new variable instead of the volume
$\mu=\ln(V/V_0)$, and the equations governing the process are
\begin{eqnarray}
\frac{dr}{d\tau}& =& 3ur(1-r^2)-r(1-r)\ ,\nonumber\\
\frac{d\mu}{d\tau}&=& -(3u+r)\ ,\nonumber\\
\frac{dn_B}{d\tau}&=& u\exp(\mu)\ ,\\
\frac{dn_M}{d\tau}&=& r\exp(\mu)\ ,\nonumber\\
\frac{dn_{\bar{B}}}{d\tau}&=& ur^3\exp(\mu)\ .\nonumber
\end{eqnarray}
In this case the system shrinks almost exponentially, when $u$ is large,
with time and hadronization
process finishes very quickly. So one can take, as an approximation, $r$ in
last two equations as a constant. Then one can see that with increase of $u$
the baryon yield can be larger than that for meson at $u\sim\kappa$.
When $u$ is large enough the anti-baryon yield can also be larger than that
of meson.
The calculated baryon and meson yields are shown in Fig. 3 for this case.
Numerical calculations confirm above estimate, as shown in Fig. 3.
The baryon to meson ratio, $R_{B/M}$, is much larger than for the first
case, as shown in Fig. 4 for both $\kappa=0.8$ and 0.6. The smaller $\kappa$ the
larger baryon to meson ratio, as argued for the first case.
\begin{figure}[tbph]
\centerline{\includegraphics[width=0.5\textwidth]{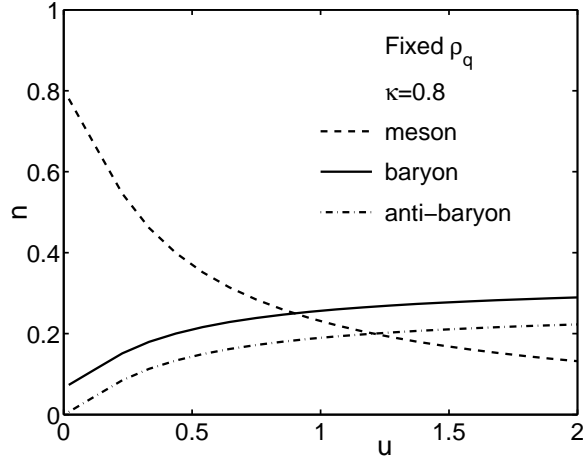}}
\caption{Yields for baryon, meson, and anti-baryon for the case with fixed quark
density as functions of $u$ for given $\kappa=0.8$.}
\end{figure}
\begin{figure}[tbph]
\centerline{\includegraphics[width=0.5\textwidth]{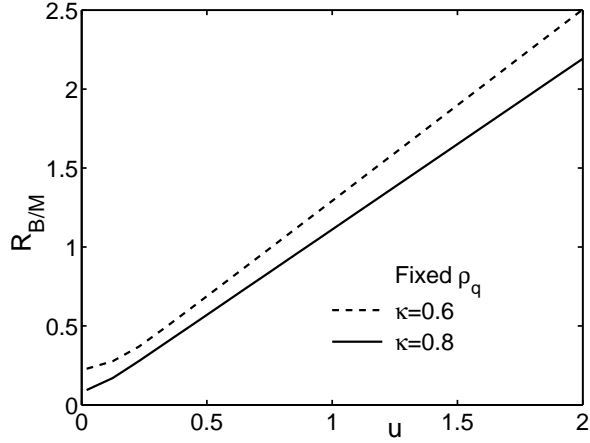}}
\caption{Ratio of baryon yield to that of meson as a function of $u$
for given $\kappa=0.8$ and 0.6 for the second case.}
\end{figure}

\subsection{For the case with hydrodynamic flow}
Now we consider a more realistic case when the partonic system is assumed to expand according
to some rules from the hydrodynamics. The true velocity profile of the produced partonic
system from the study of hydrodynamics for the evolution is quite complicated and is not
suitable for the study in this paper. We now assume that the expansion rate is proportional
to the distance from the center, as for the galaxies \cite{hydro}. If the partonic system is
regarded as an expanding ellipsoid with principal axes $a$ and $b$ then one roughly has
$da/d\tau=\nu a$ and $db/d\tau=\nu b$ with $\nu$ a constant. In this case, one can
have approximately
\begin{equation}
\frac{dV}{d\tau}=\nu V\ ,
\end{equation}
and the densities of quarks and anti-quarks decrease rapidly with $\tau$,
because of the decrease of quark number from hadronization and the increase of system's
volume. Then we define $\rho_1=\rho_q/\rho_0, \rho_2=\rho_{\bar{q}}/\rho_0$,
and get
\begin{eqnarray}
\frac{d\rho_1}{d\tau}& =& -3u\rho_1^3-\rho_1\rho_2-\nu\rho_1\ ,\nonumber\\
\frac{d\rho_2}{d\tau}&=& -3u\rho_2^3-\rho_1\rho_2-\nu\rho_2\ ,\nonumber\\
\frac{dn_B}{d\tau}&=& u\rho_1^3\exp(\nu\tau)\ ,\\
\frac{dn_M}{d\tau}&=& \rho_1\rho_2\exp(\nu\tau)\ ,\nonumber\\
\frac{dn_{\bar{B}}}{d\tau}&=& u\rho_2^3\exp(\nu\tau)\ ,\nonumber
\end{eqnarray}
with $\rho_1(0)=1, \rho_2(0)=\kappa$. We choose the parameter for system's expansion
$\nu=0.1$ in our calculation as an example. In this case the hadronization process is similar
to the first case in the beginning of hadronization with the decrease of parton densities
a little faster than in the first case. Because of the $\rho^3$ dependence of baryon production
rates, most baryons are produced when the parton density is high. So the expansion
rate $\nu$ has smaller influence on baryon production than for mesons. Consequently the
meson yield is a little bit less than in the first case. The results for the yields
are shown in Fig. 5 for $\kappa=0.8$ and the corresponding baryon to meson ratio is given
in Fig. 6 for $\kappa=0.8$ and 0.6.
\begin{figure}[tbph]
\centerline{\includegraphics[width=0.5\textwidth]{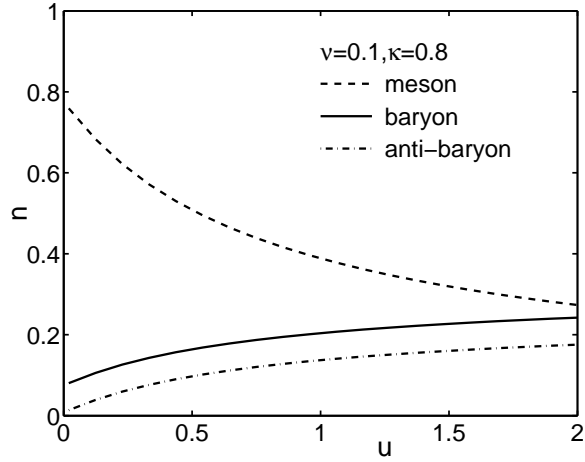}}
\caption{Yields for baryon, meson, and anti-baryon as functions of $u$ for the case
with hydrodynamical expansion for the partonic system for given $\kappa=0.8$.}
\end{figure}
\begin{figure}[tbph]
\centerline{\includegraphics[width=0.5\textwidth]{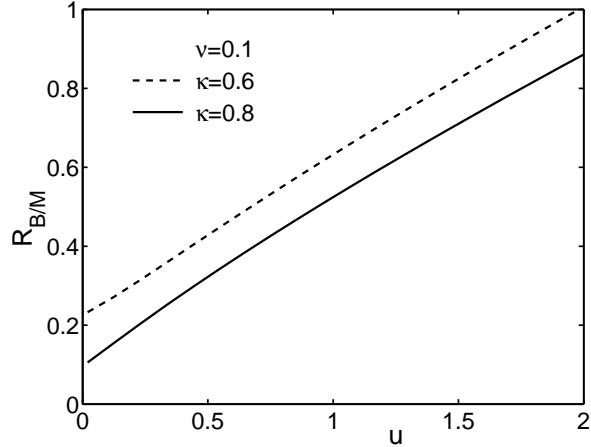}}
\caption{Ratio of baryon yield to that of meson as a function of $u$
for given $\kappa=0.8$ and 0.6 for the third case.}
\end{figure}

\section{Discussions}
From the results obtained for the three cases discussed above one can see
that the yields of baryons and mesons depend on the details of the
evolution of the system. From experimentally measured yields it is possible to
constrain the dynamics in the hadronization process. The net baryon
density influence strongly on baryon and meson yields and their ratio.
More realistic investigations on this subject, with the physical
expansion of the partonic system taken into account, are needed to draw
more reliable conclusions.

\ack
This work was supported in part by the National Natural
Science Foundation of China under Grant Nos. 10635020 and 10475032,
by the Ministry of Education of China under Grant No. 306022
and project IRT0624.


\begin{thebibliography}{0}
\bibitem{string} Braun M A, Moral F del, and Pajares C2003 \NP
{\bf A 715}, 791
\bibitem{p_pi}Adcox K et al., PHENIX Collaboration 2002 \PRL {\bf 88} 242301\\
Alder S S et al., PHENIX Collaboration 2003 \PRL {\bf 91}, 172301\\
2004 \PR {\bf C 69} 034909

\bibitem{flow}Sorensen P et al. STAR Collaboration 2004 \jpg {\bf 30} S217

\bibitem{cronin}Cronin J W et al. 1975 \PR {\bf D 11} 3105

\bibitem{asso} Adams J et al. STAR Collaboration 2005 \PRL {\bf 95} 152301\\
Adler S S et al. PHENIX Collaboration 2006 \PRL {\bf 97} 052301

\bibitem{reco} Hwa R C and Yang C B 2004 \PRL {\bf 93} 082302\\
ibid 2003 \PR {\bf C 67} 034902

\bibitem{mv} Moln\'ar D and Voloshin S A 2003 \PRL {\bf 91} 092301

\bibitem{anom}Greco V, Ko C M, and L\'evai P 2003 \PRL {\bf 90} 202302

\bibitem{fb}Fries J, M\"uller B, Nonaka C and Bass S A 2003 \PRL {\bf 90}
202303\\
 ibid 2003 \PR {\bf C 68} 044902

\bibitem{yang1}Yang C B 2006 \jpg {\bf 32} L 11

\bibitem{coals} Baltz A J et al. 1994  \PL {\bf B 325} 7

\bibitem{react} H\"anggi P, Talkner P and Borkover M 1990 \RMP {\bf 62} 251

\bibitem{hydro} Harwit M 1973 {\it Astrophysics Concepts} (Wiley, New York)\\
Misner Ch W, Thome K S and Wheeler J A 1972 {\it Gravitation} (Freeman, San Francisco)

\end{thebibliography}
\end{document}